\documentclass[prl,twocolumn,superscriptaddress,showpacs,floatfix,amsfonts]{revtex4}
\usepackage{graphicx,graphics,color,epsfig}% Include figure files
\usepackage{bm}
\usepackage{amsmath}
\usepackage{amssymb}
\usepackage{epstopdf}
\begin{document}
\preprint{}
\title{Zero-Temperature Magnetic Transition in an Easy-Axis Kondo Lattice Model}
\author{Jian-Xin Zhu}
\affiliation{Theoretical Division, Los Alamos National Laboratory,
Los Alamos, New Mexico 87545, USA}
\author{Stefan Kirchner}
\affiliation{Department of Physics \& Astronomy, Rice University,
Houston, TX 77005, USA }
\author{Ralf Bulla}
\affiliation{Theoretische Physik III, Elektronische Korrelationen
und Magnetismus, Institut f\"{u}r Physik, Universit\"{a}t Augsburg,
D-86135 Augusburg, Germany}
\author{Qimiao Si}
\affiliation{Department of Physics \& Astronomy, Rice University,
Houston, TX 77005, USA }
\begin{abstract}
We address the quantum transition of a spin-$1/2$ antiferromagnetic Kondo
lattice model with an easy-axis anisotropy using the extended dynamical
mean field theory. We derive results in real frequency using the bosonic
numerical renormalization group (bNRG) method and compare them with 
Quantum Monte Carlo results in Matsubara frequency. 
The bNRG results show a logarithmic divergence in the critical local
spin susceptibility, signaling a destruction of Kondo screening.
The $T=0$ transition is 
consistent with being
second order.
The bNRG results also display some subtle features;
we identify their origin and suggest means for further microscopic
studies.
\end{abstract}
\maketitle

\narrowtext

A sizable number of (nearly) stoichiometric
heavy fermions have recently
been discovered in which the antiferromagnetic
transition temperature can be continuously suppressed to
zero~\cite{vonLohneysen.06}.
These 
materials have not only allowed further elucidation of
the heavy fermion physics but also provided 
a concrete setting to address the
larger question
of quantum criticality. The application of the 
Landau
paradigm
considers the fluctuations of 
the magnetic order parameter as the primary critical 
modes~\cite{Hertz76}.
The resulting $T=0$ spin-density-wave (SDW) quantum
critical point (QCP)~\cite{Hertz76,Moriya85,Millis93} is Gaussian.
However, a host of dynamical, transport and thermodynamic
data~\cite{vonLohneysen.06,Schroder00,Aronson95,Paschen04,Kuchler03}
suggest that the observed QCPs
are non-Gaussian, indicating the existence of additional
quantum critical modes. Since there is not yet a universal
prescription for the identification of such emergent critical modes, 
microscopic considerations have been playing an important role.

One 
idea invokes the breakdown of the Kondo screening
effect at the magnetic QCP to characterize the new critical 
modes~\cite{Si-Nature01,PColeman01,TSenthil04}.
In 
local quantum criticality~\cite{Si-Nature01},
the destruction of the Kondo effect arises through the decoherence
by the magnetic order parameter fluctuations. Microscopically,
this picture has been studied through the extended dynamical mean
field theory (EDMFT) approach~\cite{SiSmith96,Chitra00}.
Here, the Kondo lattice systems are analyzed
in terms of a Bose-Fermi Kondo (BFK) model, with the spectra
of its fermionic and bosonic baths self-consistently determined.
The EDMFT approach
addresses
the RKKY-Kondo competition,
going beyond 
the seminal works of Refs.~\cite{Doniach77,Varma76}
in 
ways that
are important for the collapse of the Kondo scale at the
magnetic
QCP.
It 
treats 
this competition
dynamically.
Equally important,
it 
incorporates
not only paramagnetic/antiferromagnetic
phases with a ``large'' Fermi surface,
but also an
antiferromagnetic phase with a ``small'' Fermi surface 
(local moments
not participating in the electronic Fermi volume).
The critical behavior of the BFK model 
was shown to allow~\cite{Si-Nature01}
a self-consistent solution
in which the criticality of the BFK model
-- with critical Kondo screening
-- is manifested at the magnetic QCP of the lattice. 
This analytical
result was
verified in a Quantum Monte Carlo (QMC) study of a Kondo lattice model
with an easy-axis anisotropy~\cite{GrempelSi03}.
An important question is whether the actual zero-temperature 
transition is second order. Earlier works at finite temperatures,
using various QMC approaches, have led to
some conflicting 
conclusions~\cite{ZhuGrempelSi03,SunKotliar03}.
The differences have been attributed to the different EDMFT equations,
which handle the generated RKKY interactions on the ordered side
differently~\cite{SunKotliar05,SiZhuGrempel05}.

In this Letter, we study the magnetic transition of the 
anisotropic Kondo lattice model directly at zero temperature,
using the recently developed bNRG method~\cite{RBulla,MTGlossop05}.
Our results are important for experiments, not only because the
numerical studies
play an
important role in the understanding of the unusual magnetic
dynamics~\cite{Schroder00}
(which itself was the
primary initial experimental indication for the non-SDW nature of the QCP),
but also because the theoretical picture has crucial predictions for
other experiments
that are actively being examined by
on-going experiments ({\it e.g.}, Refs.~\cite{Paschen04,Kuchler03}).
More generally, whether unconventional QCPs 
would be stable and relevant to realistic models/materials or 
tend to be pre-emptied by first order transitions are broadly
important and also arises~\cite{Kuklov.04} in,
{\it e.g.}, the case of deconfined 
quantum criticality~\cite{Senthil-Science04} in spin/boson
lattice systems.

The Kondo lattice Hamiltonian is
\begin{equation}
\mathcal{H} = \sum_{ ij\sigma} t_{ij} ~c_{i\sigma}^{\dagger}
c_{j\sigma} + \sum_i J_K ~{\bf S}_{i} \cdot {\bf s}_{c,i}
+ \sum_{ ij} (I_{ij} /2) ~S_{i}^z S_{j}^z .
\label{EQ:kondo-lattice}
\end{equation}
Here, 
${\bf S}_{i}$ and ${\bf s}_{c,i}$ represent the spins of the $S={1 \over 2}$
local moment and conduction $c$-electrons respectively.
There are $1$ moment and, on average, $x<1$ conduction electrons, per site.
$J_K$ is the antiferromagnetic Kondo interaction. 
$t_{ij}$ is the hopping
integral, corresponding to a band dispersion $\epsilon_{\bf k}$
whose 
density of states (DOS)
$\rho_0(\epsilon)$
is 
featureless.
$I_{ij}$ denotes the RKKY interaction;
its Fourier transform, $I_{\bf q}$, is the most negative
at an antiferromagnetic (AF) wavevector
${\bf Q}$ ($I_{\mathbf{Q}} =-I$). 
The EDMFT approach leads to
the effective impurity action~\cite{ZhuGrempelSi03}
\begin{eqnarray}
\mathcal{S}_{\text{imp}}
&=&\mathcal{S}_{\text{top}} + \int_{0}^{\beta} d\tau
[h_{\text{loc}}~ S^{z}(\tau)
+J_K \mathbf{S}(\tau)\cdot \mathbf{s}_{c}(\tau)
] \nonumber \\
&&-\int \int_{0}^{\beta} d\tau d\tau^{\prime}
\sum_{\sigma}
c_{\sigma}^{\dagger}(\tau)G_{0,\sigma}^{-1}
(\tau-\tau^{\prime})c_{\sigma}(\tau^{\prime})
\nonumber \\
&& -\frac{1}{2}\int \int_{0}^{\beta} d\tau
d\tau^{\prime}
S^{z}(\tau)\chi_{0}^{-1}(\tau-\tau^{\prime})S^{z}(\tau^{\prime})\;.
\label{EQ:impurity-action}
\end{eqnarray}
where 
$\mathcal{S}_{\text{top}}$ is the
Berry phase of the local moment, and $h_{\text{loc}}$,
$G_{0,\sigma}^{-1}$, and $\chi_{0}^{-1}$ are the static
and
dynamical Weiss fields satisfying the 
self-consistency condition:
\begin{subequations}
\begin{eqnarray}
h_{\text{loc}}&=&-[I-\chi_{0}^{-1}(\omega=0)]~m_{\text{\tiny AF}
}\;, \label{EQ:self-consistent-a} \\
\chi_{\text{loc}} (\omega) &=&
\int_{-I}^I d \epsilon 
~{\rho_I(\epsilon) / [M(\omega)+ \epsilon]},
\label{EQ:self-consistent-b} \\[-1ex]
G_{\text{loc},\sigma} (\omega) &=&
\int_{-D}^{D} d \epsilon 
~{\rho_0(\epsilon) / [\omega + \mu -\epsilon 
- \Sigma_{\sigma}(\omega)]} \;.
\label{EQ:self-consistent-c}
\end{eqnarray}
\end{subequations}
Here, $M(\omega)$ and $\Sigma_{\sigma}(\omega)$ are respectively
the spin and conduction-electron self-energies, which
satisfy the Dyson(-like) equations:
$\Sigma_{\sigma}(\omega)=G_{0,\sigma}^{-1}(\omega) 
-G_{\text{loc},\sigma}^{-1}(\omega)$, and 
$ M(\omega)=
\chi_{0}^{-1}(\omega)+\chi_{\text{loc}}^{-1}(\omega)$. 
$m_{\text{\tiny AF}}=\langle S^{z}\rangle_{\text{imp}}$ is the staggered
magnetization; $\chi_{\text{loc}} (\omega)$ and
$G_{\text{loc},\sigma} (\omega)$ are
the connected local spin susceptibility and local
conduction electron Green's function,
respectively. Finally,
$M(\omega)$ also specifies the lattice spin 
susceptibility~\cite{SiSmith96}:
\begin{eqnarray}
\chi({\bf q},\omega) = 1/[I_{\bf q} + M(\omega)]\;.
\label{chi-q-omega}
\end{eqnarray}
As described in detail
in Refs.~\cite{GrempelSi03,ZhuGrempelSi03},
the effective impurity action [Eq.~(\ref{EQ:impurity-action})]
can be rewritten in a Hamiltonian form,
in which the dynamical Weiss fields are represented 
by a fermionic bath and a bosonic one. 
Through a canonical transformation,
the fermionic coupling
is reduced to a transverse field
Ising model with
an ohmic bosonic bath.
Integrating out the two bosonic
baths yields a form that is suitable for QMC studies:
\begin{eqnarray}
{\cal S}_{\text{imp}}'=&& \int_0^{\beta} d \tau [ h_{\text{loc}}
~S^z (\tau) + \Gamma S^x(\tau) -{1 \over 2} \int_0^{\beta} d \tau'
S^z(\tau) 
%2col
\nonumber\\
&&\times S^z(\tau')
(\chi_0^{-1}(\tau-\tau') - {\cal K}_c(\tau-\tau') ) ] .
\label{Z-imp}
\end{eqnarray}
Here, ${\cal K}_c (i\omega_n) 
= \kappa_c |\omega_n| $ describes
the ohmic dissipation;
$\kappa_c$ and $\Gamma$ 
are determined by the longitudinal and transverse components
of the Kondo coupling, respectively.

For the bNRG studies, we work in the real frequency domain by rewriting
Eq.~(\ref{Z-imp}) in a Hamiltonian form:
\begin{eqnarray} 
{\mathcal{H}}_{\text{imp}}'
 &=& h_{\text{loc}} ~S^z + \Gamma S^x 
%2col
\nonumber\\ 
&&
%2col
+ \;  
 \sum_{p} \tilde{g}_{p} S^z \left( \phi_{p} +
\phi_{-p}^{\;\dagger} \right) + \sum_{p}
\tilde{w}_{p}\,\phi_{p}^{\;\dagger} {\phi}_{p}\;,
\label{EQ:Heff-imp'}
\end{eqnarray}
where $\tilde{\omega}_{p}$ and $\tilde{g}_{p}$ are such that
$-\sum_{p}\frac{2\tilde{g}_{p}^{2} \tilde{\omega}_{p}}{\omega^{2} -
\tilde{\omega}_{p}^{2}} 
= \tilde{\chi}_{0}^{-1}(\omega)
\equiv \chi_{0}^{-1}(\omega) - \mathcal{K}_{c}(\omega)$.
The EDMFT 
procedure starts with 
a trial $h_{\text{loc}}$ and 
$\chi_{0}(\omega)$.
The bNRG iteration loop~\cite{RBulla}
is then used to solve the impurity model
(\ref{EQ:Heff-imp'})
for 
$m_{\text{\tiny AF}}$ and 
$ \chi_{\text{loc}} (\omega)$ which,
in turn, lead to updated 
$h_{\text{loc}}$ and $\chi_{0}^{-1}(\omega)$.   
The procedure is repeated until 
convergence is achieved.
For the most part, we consider two-dimensional magnetic fluctuations~\cite{ZhuGrempelSi03} 
as represented by a constant RKKY DOS
%\begin{equation}
$\rho_{I} (\epsilon) \equiv  \sum_{\bf q} \delta ( \epsilon  -
I_{\bf q} ) = (1/{2I})\Theta(I - | \epsilon | ) $,
%\label{EQ:rkky-dos}
%\end{equation}
with $\Theta$ being the Heaviside function.  
In this case,
Eq.~(\ref{EQ:self-consistent-b})
yields
\begin{equation}
M(\omega)=I/\tanh[I\chi_{\text{loc}}(\omega)]\;.
\label{EQ:Self-spin}
\end{equation}
We take the energy cutoff $\omega_{\text{cutoff}}=1$ and
the parameters $\Gamma=0.75$ and $\kappa_c=\pi$,
yielding
$T_K^0 \equiv 1/\chi_{\text{loc}}(\omega=0,I=0) \approx 0.71$.
In most cases (exceptions will be specified),
we choose the NRG discretization parameter
$\Lambda=2$, 
keep
$N_{b}=100$ bosonic states 
for 
the impurity site and 8 states
for the other sites,
and retain $N_s=60$ many-body states. 
To reach convergence,
the difference 
between two consecutive iterations 
in 
$h_{\text{loc}}$ as well as in $\chi_0^{-1}(\omega)$ for each $\omega$
is smaller than $10^{-6}$.
Away from the transition region, 30 or so EDMFT iterations are sufficient.
In the transition region, it takes as many as 2400 iterations.
%The large number of EDMFT iterations needed,
%together with the bNRG iterations for each EFMFT iteration,
%dictate
%the numerically 
%intensive nature of our study.

\begin{figure}[t]
\centerline{\psfig{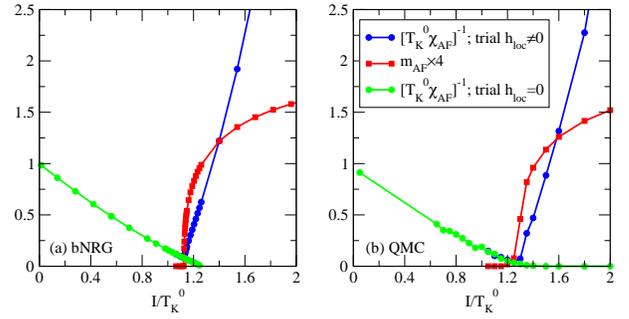}}
\caption{(Color online) Inverse static AF susceptibility,
$\chi_{\text{\tiny AF}}^{-1} \equiv \chi({\bf q= Q}, \omega=0)^{-1}
=M(\omega=0)-I$,
from the PM (trial $h_{\text{loc}}=0$; green circles)
and AF (trial $h_{\text{loc}} \ne 0$; blue circles) solutions and the
AF order parameter $m_{\text{\tiny AF}}$ (red squares), obtained from the
bNRG (a) and QMC (b) methods. The lines are guides to eyes.
See the main text for details.
}
\label{FIG:PhaseDiagram}
\end{figure}

The resulting phase diagram is summarized in Fig.~\ref{FIG:PhaseDiagram}(a). 
We observe a substantial drop of $\chi_{\text{\tiny AF}}^{-1}$
from both sides, as well as of $m_{\text{\tiny AF}}$. 
Indeed, the magnetic order parameter 
$m_{\text{\tiny AF}}$
vanishes
continuously within the numerical uncertainty
as $I$ approaches the transition point $I_{c1}$ 
($\approx 1.1228T_K^0$).

\begin{figure}[t]
\centerline{\psfig{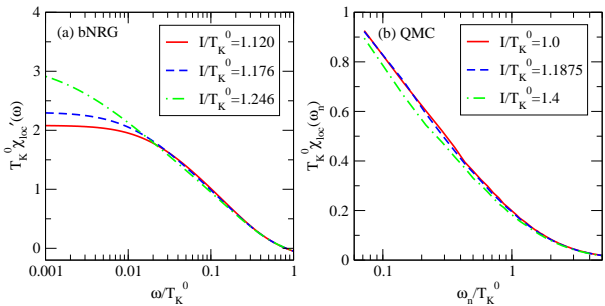}}
\caption{(Color online) Frequency dependence of the local spin
susceptibility
at various values of $I$ around the magnetic transition.
(a) $\chi^{\prime}_{\text{loc}}(\omega)$ vs. the real frequency $\omega$,
from bNRG; (b) $\chi_{\text{loc}}(\omega_n)$ vs. the Matsubara
frequency $\omega_n$, from QMC.
}
\label{FIG:LocalSuscep}
\end{figure}

Fig.~\ref{FIG:LocalSuscep}(a) shows $\chi_{\text{loc}}(\omega)$
at various $I$,
from around $I=I_{c1}$ and beyond.
Above a cutoff scale,
the local susceptibility is logarithmically dependent on the frequency.
Such a singular behavior signals the Kondo screening being critical,
which is the hallmark of 
local quantum criticality. Fitting the slope of the logarithmic dependence
in terms of $\alpha/2I$ yields an $\alpha$ which is nearly constant
(varying by less than 2\%) in the shown range of $I$.
Through the self-consistency
 Eq.~(\ref{EQ:Self-spin}) together with Eq.~(\ref{chi-q-omega}),
$\alpha$ is the critical exponent
that appears in the dynamical AF spin susceptibility,
$\chi_{\text{\tiny AF}}(\omega)
\equiv
\chi({\bf Q},\omega)$.
Extrapolating to the 
bNRG continuum limit ($\Lambda \rightarrow 1^+$)
yields
$\alpha \approx 0.83$.
The low-frequency cut-off scale for the logarithmic dependence
is relatively small, becoming of the order of 
$\sim 10^{-2} T_K^0$ for the largest $I$ we have reached
as $I$ is increased towards $I_{c2}$ ($\approx 1.26 T_K^0$),
the instability point of the paramagnetic solution signaled
by a diverging $\chi_{\text{\tiny AF}}$;
this cut-off scale 
extrapolates to zero as $I \rightarrow I_{c2}^-$.

For comparison, the QMC results for the phase diagram and the Matsubara
frequency dependence of the local dynamical spin susceptibility
are shown in Figs.~\ref{FIG:PhaseDiagram}(b) and 
\ref{FIG:LocalSuscep}(b), respectively. At the gross level, the bNRG 
and QMC results are 
similar to each other. 

At a fine level, the bNRG results contain some 
differences
from their QMC counterparts.
Major among these is the observation,
as seen in Fig.~\ref{FIG:PhaseDiagram}(a), 
that $I_{c2}$ 
is larger than $I_{c1}$ by about 12\%.
In the QMC results, by contrast, $I_{c2}$ equals $I_{c1}$ within
the numerical uncertainty of a few percents. To see whether this is unique
to the bNRG results for the 2D magnetic fluctuations, we have carried
out similar bNRG studies of the EDMFT phase diagram in the case of 
3D magnetic fluctuations 
-- as represented by a semicircular RKKY DOS, 
$\rho_I(\epsilon)=\frac{2}{\pi I^{2}}\sqrt{I^{2}-\epsilon^{2}}
\theta(I-\vert \epsilon\vert)$. 
The 3D case does not have the complication of a divergent
local susceptibility, and a SDW solution is expected
in the EDMFT approach~\cite{Si-Nature01,GrempelSi03,ZhuGrempelSi03}.
We find that the magnetic transition is essentially continuous
(with the upper bound of the order-parameter jump being 0.016),
yet $(I_{c2}-I_{c1})/I_{c1}$ is still non-zero (about $13\%$).

The observation of a continuous
onset of the magnetic order parameter 
$m_{\text{\tiny AF}}$ but,
at the same time, 
different $I_{c1}$ and $I_{c2}$,
is unexpected.  
One possibility is that 
the dichotomy 
is inherent to the
EDMFT equations.
To address this, we 
return to  the self-consistent equation
for the magnetic order parameter,
Eq.~(\ref{EQ:self-consistent-a}). For 
a small
$h_{\text{loc}}$
(we have numerically determined that the magnetic solution
is the same regardless of whether an infinitesimal finite
value or a large value is chosen for the trial $h_{\text{loc}}$),
we have $m_{\text{\tiny AF}} = - \chi_{\text{loc}} h_{\text{loc}} 
- a_3 h_{\text{loc}}^3 
- a_5 h_{\text{loc}}^5 + \dots$. Note,
the linear coefficient is equal to $-\chi_{\text{loc}}(\omega=0)$
since,
in Eq.~(\ref{EQ:Heff-imp'}), 
$h_{\text{loc}}$ couples linearly to $S^z$ only~\cite{note-quadratic}.
We can then rewrite
Eq.~(\ref{EQ:self-consistent-a})
as 
\begin{equation}
r h_{\text{loc}}= -u h_{\text{loc}}^3 - v h_{\text{loc}}^5 + \dots \;,
\label{EQ:landau}
\end{equation}
where
$r = \chi_{\text{loc}}/\chi_{\text{\tiny AF}}$
is the quadratic coefficient of the 
corresponding static Landau function,
and 
%\begin{equation}
%%$r=\chi_{\text{loc}}/\chi_{\text{\tiny AF}}$,
$u=-a_{3}[\chi_{\text{loc}}^{-1} - \chi_{\text{\tiny AF}}^{-1}]$
and $v=-a_{5}[\chi_{\text{loc}}^{-1} - \chi_{\text{\tiny AF}}^{-1}]$
are 
the quartic and sextic Landau coefficients.
When $u>0$
(the alternative, $u<0$, would lead to a large jump
in $m_{\text{\tiny AF}}$, in contrast to what
we have observed),
we have a canonical case of a second-order 
transition
at $r=0$
(in other words,
a $h_{\text{loc}} \ne 0$ solution cannot occur for any $r>0$).
Through $r = \chi_{\text{loc}}/\chi_{\text{\tiny AF}}$,
this implies that, at $I_{c1}$
(the onset of the magnetic transition)
$\chi_{\text{\tiny AF}}$ diverges.
This is the same condition for
$I_{c2}$, where the paramagnetic solution goes away.
So, within the EDMFT equations per se,
a continuous onset in $m_{\text{\tiny AF}}$ must
coincide with a vanishing $(I_{c2}-I_{c1})$.

\begin{figure}[t]
\centerline{\psfig{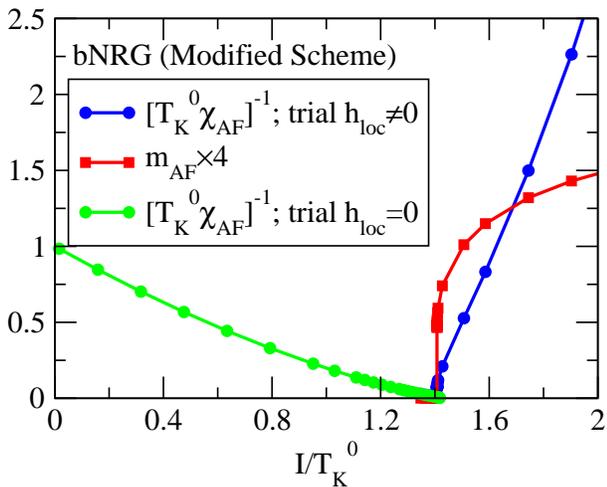}}
%\centerline{\psfig{figure=nfig3.eps,height=8cm,width=8cm,angle=0}}
\caption{(Color online) Inverse AF susceptibilities
(PM, green circles; AF, blue circles) and
the AF ordered moment (red squares), from
the modified scheme, as described
in the main text. The notations are the same as
in Fig.~\ref{FIG:PhaseDiagram}.
}
\label{FIG:ModifiedSchemes}
\end{figure}

We are then led to search for numerical origins for the 
dichotomatic observation, and have identified the primary
source.
Within bNRG,
as in any NRG method,
the imaginary part of the local susceptibility, 
$\chi^{\prime\prime}_{\text{loc}}(\omega)$,
is calculated in terms of a set of Gaussian-broadened delta functions.
The real part is in turn determined via the Kramers-Kronig relation,
which we call $\chi^{\prime}_{\text{loc},\text{KK}}(\omega)$.
The static local susceptibility can alternatively be calculated in terms 
of a) the differential response of the local magnetization {\em w.r.t.}
$h_{\text{loc}}$ or b) $\sum_{n}| \langle n|S_z|0\rangle |^2/(E_n-E_0)$,
where $n$ labels all the 
many-body excited 
states and $0$ the ground state. We find that the latter two methods
yield essentially the same result, which we call 
$\chi_{\text{loc},\text{static}}$.
A key observation is that $\chi_{\text{loc},\text{static}}$ 
is larger than $\chi^{\prime}_{\text{loc},\text{KK}}(\omega=0)$
by a sizable amount (about 11.5\% for $\Lambda$=2, in the 2D case). 
The quadratic Landau coefficient then becomes
$
%\begin{equation}
r = \chi_{\text{loc,static}}
/\chi_{\text{\tiny AF}}
-
[\chi_{\text{loc,static}}/
\chi_{\text{loc,KK}}(\omega=0) - 1 ] .
\label{EQ:landau_quadratic2}
%\end{equation}
$
It follows that the onset of the magnetic transition (at $r=0$)
already occurs before $\chi_{\text{\tiny AF}}$
diverges, which explains the $I_{c1} < I_{c2}$ discussed earlier.
In order to confirm our observation, we 
have implemented 
the simplest modification scheme
to ensure that the Kramers-Kronig of the 
NRG-calculated $\chi^{\prime\prime}_{\text{loc}}(\omega)$
yields a static local susceptibility that is equal to
$\chi_{\text{loc},\text{static}}$.
We use, during each EDMFT iteration, 
$\chi_{\text{loc},\text{static}}$ for the $\omega=0$ 
component
of $\chi^{\prime}_{\text{loc}}(\omega)$,
but retain
$\chi^{\prime}_{\text{loc},\text{KK}}(\omega)$
for
all finite frequencies.
%%The numerical effort is even larger,
%%requiring more than 5000 EDMFT iterations
%%near the transition.
We find that $I_{c1}$ is increased 
compared
to that of the ``vanilla'' scheme.
($I_{c2}$ is essentially unchanged, although 
the normalization parameter $T_K^0$ is reduced.)
Moreover, as shown in Fig.~\ref{FIG:ModifiedSchemes}
for the 2D case,
$I_{c2} \approx I_{c1}$ [with a difference
less than 1\% (2\%) in the 2D (3D) case,
with 
$\Lambda=2$].
$m_{AF}$ vs. $(I-I_{c1})$ from 
the modified scheme  is mostly comparable 
to that of Fig.~\ref{FIG:PhaseDiagram}(a) except for
being steeper in the immediate vicinity of $I_{c1}$
[when $(I-I_{c1})/I_{c1}$ is within a few percents].
The magnetic transition is therefore 
second order
within the numerical accuracy.

%Since $I_{c2}$ is
%%% by definition 
%where the 
%AF susceptibility diverges, 
%a coinciding $I_{c1}$ and $I_{c2}$
%implies that the AF susceptibility diverges 
%at the magnetic transition. This is clearly
%seen in the plots of the inverse AF susceptibility
%given in Fig.~\ref{FIG:ModifiedSchemes} [as well as 
%at $I_{c2}$ in Fig.~\ref{FIG:PhaseDiagram}(a)].
%%The magnetic transition is therefore 
%%second order
%%within the numerical accuracy.
%%For the 2D case,
%%the divergent AF susceptibility
%%implies a divergent local susceptibility
%%[as 
%%%%can be 
%%seen from 
%%Eqs.~(\ref{EQ:self-consistent-b},\ref{chi-q-omega}),
%%using a constant 
%%$\rho_I$].

For 
$I>I_{c2}$~\cite{note-inst},
the nominally self-consistent 
paramagnetic
solution has ${\rm max}[2I\chi_{\text{loc}}^{\prime\prime}(\omega)] 
> \pi$, which, through 
Eq.~(\ref{EQ:Self-spin}),
yields an oscillatory 
$M^{\prime\prime}(\omega)$~\cite{Haule.03}.
By contrast, in the Matsubara frequency domain, a nominally
paramagnetic solution 
($h_{\text{loc}}=0=m_{\text{\tiny AF}}$, but a finite Curie constant)
exists
%% even 
%%inside the magnetic part of the phase diagram
for $I>I_{c2}$,
%%~\cite{ZhuGrempelSi03},
which helped to determine
the
phase diagram~\cite{ZhuGrempelSi03}. 
%It would be helpful to carry out
%QMC calculations
%at very low temperatures;
%the cluster-MC 
%method~\cite{Niedermayer.88}, which may
%reach
%$T \sim 10^{-5}T_K^0$, is a promising route to pursue.

Independently, Glossop and Ingersent~\cite{GlossopIngersent}
have 
carried out
NRG studies within the EDMFT approach to the same Kondo
lattice model. They
used
the NRG method of Ref.~\cite{MTGlossop05},
in which the 
Kondo coupling to the conduction electrons are directly
treated (instead of being mapped to an Ohmic dissipation).
%(By contrast, we have mapped the Kondo coupling 
%to an ohmic dissipative bath and applied the bNRG method
%of Ref.~\cite{RBulla}.)
Moreover, in Ref.~(\cite{GlossopIngersent}b), they
adopted a somewhat different modification
scheme to ensure the consistency between the static
local susceptibilities from two ways of calculation within NRG.
In spite of these differences in methods, 
the results from the two groups are largely compatible
with each other.

To summarize, we have carried out bosonic numerical 
renormalization group studies of the extended dynamical
mean field theory of a Kondo lattice model.
The local spin susceptibility 
has 
a logarithmic frequency dependence 
-- signifying the critical 
Kondo screening -- and 
the magnetic transition
is consistent with being second order.
These
results provide evidence for local
quantum criticality.
% and hence are important
%for experiments in heavy fermions and 
%quantum phase transitions 
%in general.
%Finally, 
Our study has also advanced the understanding
of the numerical renormalization group,
a venerable
method~\cite{Bulla-RMP}
in the area of correlated systems.

{\bf Acknowledgments:} 
We thank M. T. Glossop, K. Ingersent, G. Kotliar, 
Z. Nussinov, G. Ortiz, P. Sun, N.-H. Tong, and L. Zhu 
for useful discussions. 
We dedicate this work to the memory of the late Daniel R. Grempel,
with whom two of us (JXZ and QS) had collaborated on QMC studies.
We acknowledge the support of
%%the 
%%U.S. 
DOE 
at Los Alamos
%%LANL
under Contract No. DE-AC52-06NA25396 and Grant Nos. LDRD-DR X9GT
and JACC/CCPM/WPR0 (JXZ), NSF Grant No. DMR-0706625 and 
the Robert A. Welch Foundation (SK and QS), and the DFG
collaborative research center SFB 484 (RB).


\begin{thebibliography}{99}
\bibitem{note-inst}
The $I$ beyond which nonanalyticity develops differs
from the extrapolated $I_{c2}$ by
about 0.5\% (for $\Lambda=2$), which we 
attribute to numerical uncertainty.
\bibitem{note-quadratic}
This is equivalent to saying~\cite{SiZhuGrempel05}
that the effective RKKY interaction generated by the conduction
electrons is excluded on the ordered side, as it inherently
is on the paramagnetic side. Otherwise, a particle-hole bubble
would contribute an additional term to the quadratic Landau
coefficient on the ordered side [the parameter $r$ in
Eq.~(\ref{EQ:landau})]~\cite{SunKotliar05}, leading to
a strongly first order transition: indeed,
$(I_{c2}-I_{c1})/I_{c1}$ is larger than 100\% in Ref.~\cite{SunKotliar03}.
\end{thebibliography}
\end{document}